\documentclass[a4paper]{article}
\usepackage{amssymb}
\usepackage{graphicx}
\usepackage{algorithm}
\usepackage{algorithmic}
\usepackage{diagbox}
\usepackage{color}
\usepackage{amsmath}
\usepackage{multirow}
\usepackage{array}

\usepackage{INTERSPEECH2020}

\title{EigenEmo: Spectral Utterance Representation Using Dynamic Mode Decomposition for Speech Emotion Classification}
\name{Shuiyang Mao, P. C. Ching, Tan Lee}
\address{Department of Electronic Engineering, The Chinese University of Hong Kong, Hong Kong}
\email{maoshuiyang@link.cuhk.edu.hk, pcching@ee.cuhk.edu.hk, tanlee@ee.cuhk.edu.hk}

\begin{document}

\maketitle
\begin{abstract}
Human emotional speech is, by its very nature, a variant signal. This results in dynamics intrinsic to automatic emotion classification based on speech. In this work, we explore a spectral decomposition method stemming from fluid-dynamics, known as Dynamic Mode Decomposition (DMD), to computationally represent and analyze the global utterance-level dynamics of emotional speech. Specifically, segment-level emotion-specific representations are first learned through an Emotion Distillation process. This forms a multi-dimensional signal of emotion flow for each utterance, called Emotion Profiles (EPs). The DMD algorithm is then applied to the resultant EPs to capture the eigenfrequencies, and hence the fundamental transition dynamics of the emotion flow. Evaluation experiments using the proposed approach, which we call EigenEmo, show promising results. Moreover, due to the positive combination of their complementary properties, concatenating the utterance representations generated by EigenEmo with simple EPs averaging yields noticeable gains.
\end{abstract}

\section{Introduction}
Human emotional speech is dynamic, which, when taken into account, may augment our understanding of these complex signals and lead to modeling advancements. In the past few decades, efforts have been made to perform dynamic modeling for automatic speech emotion classification explicitly. Dynamic models, such as hidden Markov models (HMMs) \cite{schuller2003hidden, vlasenko2014modeling, vlasenko2011emotion} and recurrent neural networks (RNNs), e.\,g., with long short-memory (LSTM) \cite{wollmer2012analyzing, lee2015high, huang2016attention, mirsamadi2017automatic, han2018towards}, are frequently used. For their input features, the common practice considers frame-based low-level features such as Mel Frequency Cepstrum Coefficients (MFCCs), energy, or pitch. The final assignment of an emotion label is then based on the \textit{low-level feature fluctuations} captured by the dynamic models.

As a complement to most of the work as mentioned above, this work aims at utilizing spectral methods for the dynamic modeling of emotion. Spectral analysis is widely used in signal processing to decompose a signal into its component frequencies, thereby revealing the dominant dynamics that make up the signal and summarizing its transitions. In particular, this paper presents the Dynamic Mode Decomposition (DMD) \cite{schmid2010dynamic, kutz2016dynamic, kayal2019eigensent} algorithm to identify the dominant behavior that underlies emotional speech. The DMD algorithm was invented by P. Schmid as a diagnostic tool for extracting dynamic information from temporal measurements of a multivariate ﬂuid ﬂow vector. The dynamic modes extracted are the non-orthogonal eigenvectors of a non-normal matrix that best characterizes the one-step evolution of the measured vector \cite{prasadan2019time}, allowing for the data-driven discovery of fundamental transition dynamics. The development of DMD is timely due to the concurrent rise of data science, encompassing a broad range of techniques, from machine learning and statistical regression to computer vision and compressed sensing \cite{kutz2016dynamic}. To the best of the authors' knowledge, this is the first attempt to apply the DMD algorithm on emotional speech.
\begin{figure}[tp]
  \centering
  \includegraphics[width=\linewidth]{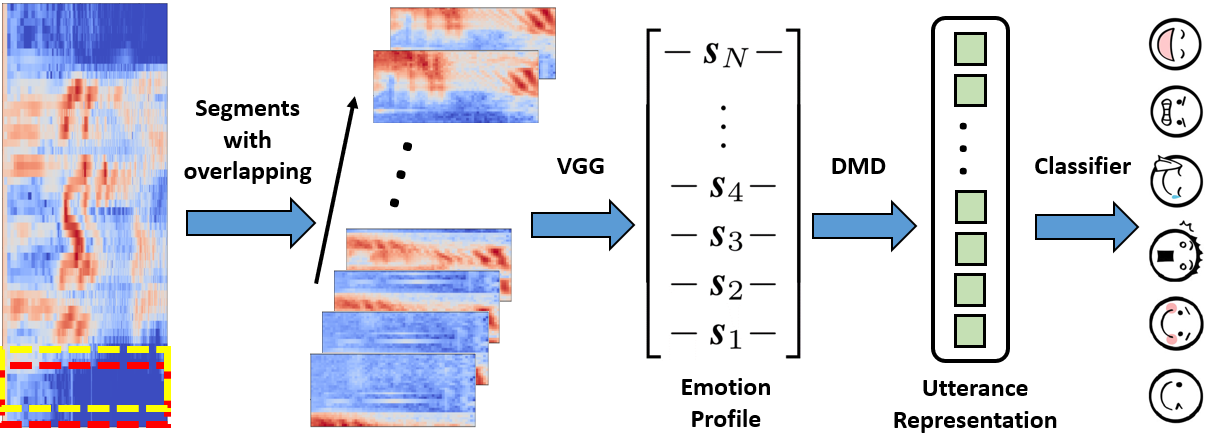}
  \caption{Illustration of the proposed method}
  \label{fig:framework}
\end{figure}

Our method builds on the concept termed Emotion Distillation, which is the process of generating a set of emotion-specific features from the original high-dimensional feature space that explicitly describes the \textit{emotion fluctuations} over time \cite{provost2012simplifying}. Distillation features are strongly tied to the task of interest, naturally highlighting salient portions of the data. In this work, we distill emotion information using Emotion Profiles (EPs) \cite{provost2012simplifying}. EPs are typically represented by a multi-dimensional signal, where each dimension represents a classifier-derived estimate of probability distribution of a set of basic emotion content (e.\,g., angry, fear, happy, sad). A large body of earlier works have demonstrated the efficacy of EPs in emotion-related tasks \cite{provost2012simplifying, mower2010framework, shangguan2015emoshapelets, kim2013emotion}. This paper extends EPs into an end-to-end approach, where EPs are learned from log Mel spectrograms via a deep Convolutional Neural Network model. Furthermore, in contrast to the aforementioned earlier works, which only utilize the final estimates to constitute their EPs, this paper also investigates the bottleneck features to form EPs. Extensive experiments are conducted on two popular emotion corpora, namely, the CASIA corpus \cite{tao2008design} and the SAVEE database \cite{jackson2014surrey}. Empirical results show the efficacy of the proposed method.

\section{Methodology}
\label{sec:majhead}

Figure~\ref{fig:framework} illustrates the proposed framework. It comprises a VGG \cite{simonyan2014very} deep Convolutional Neural Network (CNN) trained on log Mel filterbanks of individual segments to make the segment-level decision. The Emotion Profiles (EPs) are then generated and utilized for constructing utterance representations using the Dynamic Mode Decomposition (DMD) algorithm. Finally, a relatively simple classifier is employed to make the final decision.
 
\subsection{Emotion profiles (EPs)}
 Emotion Profiles (EPs) were introduced and demonstrated to be useful for emotion classification tasks in \cite{provost2012simplifying, mower2010framework, shangguan2015emoshapelets, kim2013emotion}. Typically, EPs are time series estimates of a set of the typical “basic” emotions (e.\,g., angry, happy, neutral, sad), with each EP component estimating the degree of confidence in the presence or absence of the corresponding emotion cues across the utterance. We call this kind of EPs the \textit{estimate-level} EPs (EEPs). In addition to EEPs, this work also explores the possibility of using bottleneck features for constructing EPs, called the \textit{bottleneck feature-level} EPs (BEPs) in this paper. Many works \cite{lee2017multi, yu2018multi, meng2017multi} have shown that bottleneck features contain rich information. We herein posit that the BEPs might serve as a complementary feature source to the conventional EEPs.

\subsubsection{Generating EPs}
The EPs in this work are generated using a VGG model, which is trained on the 64-bin log Mel filterbanks of individual segments. The log Mel filterbanks are computed by \textit{short-time Fourier transform} (STFT) with a window length of 25 ms, hop length of 10 ms, and FFT length of 512. Subsequently, 64-bin log Mel filterbank features are derived from each short-time frame, and the frame-level features are combined to form a time-frequency matrix representation of the segment. Each segment inherits the label of the utterance where it lies. 

The trained VGG model aims to predict a probability distribution \(\textbf{\textit{P}}_i\) for the \(i^{\textnormal{th}}\) segment in a certain utterance:
\begin{align}
  \textbf{\textit{P}}_i = [p_i(e_1),\ p_i(e_2),\ \cdots,\ p_i(e_C)]^T
  \label{eq1}
\end{align}
where, \(e_{1},\ e_{2},\ \cdots,\ e_{C}\), represent the set of basic emotions. The EEP for a specific utterance \(\textbf{\textit{U}}\) can then be expressed as
\begin{align}
  \textbf{\textit{U}}_{EEP} = [\textbf{\textit{P}}_1,\ \textbf{\textit{P}}_2,\ \cdots,\ \textbf{\textit{P}}_{N}]
  \label{eq2}
\end{align}
Where \(N\) is the number of segments in the utterance.

Meanwhile, the outputs of the penultimate layer of the trained VGG, i.\,e., the \textit{bottleneck features}, are utilized to construct the BEP for Utterance \(\textbf{\textit{U}}\) as
\begin{align}
  \textbf{\textit{U}}_{BEP} = [\textbf{\textit{B}}_1,\ \textbf{\textit{B}}_2,\ \cdots,\ \textbf{\textit{B}}_{N}]
  \label{eq3}
\end{align}
Where each \(\textbf{\textit{B}}_{i} \in \mathbb{R}^{M\times 1}, i=1,2,..,N\), represents the bottleneck feature vector for the \(i^{\textnormal{th}}\) segment in Utterance \(\textbf{\textit{U}}\), with embedding dimension of \(M\). 

\subsection{Dynamic Mode Decomposition (DMD)}

For the purposes of applying the DMD method, the following matrix is first defined:
\begin{align}
  \textbf{\textit{U}}^{k}_{j} = [\textbf{\textit{s}}_{j},\ \textbf{\textit{s}}_{j+1},\ \dots,\ \textbf{\textit{s}}_{k}]
  \label{eq4}
\end{align}
This matrix includes Segment \(j\) through \(k\) of Utterance \(\textbf{\textit{U}}\). A segment, \(\textbf{\textit{s}}_{i}\), can be replaced by a probability distribution \(\textbf{\textit{P}}_i\in \mathbb{R}^{C\times 1}\) (for EEPs), or by a bottleneck feature vector \(\textbf{\textit{B}}_{i} \in \mathbb{R}^{M\times 1}\) (for BEPs).

To construct the Koopman operator \cite{kutz2016dynamic, koopman1931hamiltonian} that best represents the data collected, the matrix \(\textbf{\textit{U}}^{N}_{1}\) (i.\,e., the whole utterance) is considered:
\begin{align}
  \textbf{\textit{U}}^{N}_{1} = [\textbf{\textit{s}}_{1},\ \textbf{\textit{s}}_{2},\ \dots,\ \textbf{\textit{s}}_{N}]
  \label{eq5}
\end{align}
Where \(N\) is the number of segments in the utterance. 

To apply standard DMD \cite{schmid2010dynamic}, the first-order Koopman assumption is employed:
\begin{equation} 
\begin{aligned}
\textbf{\textit{s}}_{k} = \textbf{\textit{A}}\textbf{\textit{s}}_{k-1}
\label{eq6}
\end{aligned}
\end{equation}
The matrix \(\textbf{\textit{U}}^{N}_{1}\) then reduces to
\begin{align}
  \textbf{\textit{U}}^{N}_{1} = [\textbf{\textit{s}}_{1},\ \textbf{\textit{As}}_{1},\ \dots,\ \textbf{\textit{A}}^{N-1}\textbf{\textit{s}}_{1}]
  \label{eq7}
\end{align}
or
\begin{align}
  \textbf{\textit{U}}^{N}_{2} = \textbf{\textit{AU}}^{N-1}_{1}
  \label{eq8}
\end{align}
Where \(\textbf{\textit{A}}\) is the Koopman operator and is chosen to minimize the Frobenius norm of \(||\textbf{\textit{U}}^{N}_{2} - \textbf{\textit{A}} \textbf{\textit{U}}^{N-1}_{1}||_F\). In other words, the operator \(\textbf{\textit{A}}\) advances each segment in \(\textbf{\textit{U}}^{N-1}_{1}\) a single time step, resulting in the corresponding future segments in \(\textbf{\textit{U}}^{N}_{2}\). The operator \(\textbf{\textit{A}}\) thus captures the overall transition dynamics of the utterance, and summarizing \(\textbf{\textit{A}}\) would lead to the construction of the desired utterance representation.

The first-order Koopman assumption constrains a segment in an utterance to transition solely from the previous one. To make our assumption more realistic, we look towards the higher-order Koopman assumption \cite{le2017higher}:
\begin{equation}
\begin{aligned}
\textbf{\textit{s}}_{k} = \textbf{\textit{A}}_{1}\textbf{\textit{s}}_{k-1} + \cdots + \textbf{\textit{A}}_{d-1}\textbf{\textit{s}}_{k-d+1} + \textbf{\textit{A}}_{d}\textbf{\textit{s}}_{k-d}
 \label{eq9}
\end{aligned}
\end{equation}
Where \(d\) is the \textit{order} parameter. This can be written in a form similar to Equation (5) and Equation (8), respectively:
\begin{equation} 
\begin{aligned}
\tilde{\textbf{\textit{U}}}^{N}_{1} = [\tilde{\textbf{\textit{s}}}_{1},\ \tilde{\textbf{\textit{s}}}_{2},\ \dots,\ \tilde{\textbf{\textit{s}}}_{N}]
\label{eq10}
\end{aligned}
\end{equation}
~\\[-0.8cm]
\begin{equation} 
\begin{aligned}
\tilde{\textbf{\textit{U}}}^{N}_{2} = \tilde{\textbf{\textit{A}}}\tilde{\textbf{\textit{U}}}^{N-1}_{1}
\label{eq11}
\end{aligned}
\end{equation}
where,
\begin{equation}
\begin{aligned}
\tilde{\textbf{\textit{s}}}_{k} = [\textbf{\textit{s}}_{k},\ \textbf{\textit{s}}_{k+1},\  \cdots,\  \textbf{\textit{s}}_{k+d-2},\ \textbf{\textit{s}}_{k+d-1}]^{T}
\label{eq12}
\end{aligned}
\end{equation}
\begin{equation} 
\begin{aligned}
\tilde{\textbf{\textit{A}}} = 
\left[
\begin{matrix}
\textbf{0}\quad\;\;\;\; \textbf{\textit{I}}\quad\;\;\;\; \textbf{0}\quad\quad \cdots\quad \textbf{0}\quad\;\;\;\; \textbf{0}\\
\textbf{0}\quad\;\;\;\; \textbf{0}\quad\;\;\;\; \textbf{\textit{I}}\quad\quad \cdots\quad \textbf{0}\quad\;\;\;\;  \textbf{0}\\
\vdots\quad\;\;\;\;\; \vdots\quad\;\;\;\;\; \vdots\quad\quad\, \ddots\quad \vdots\quad\;\;\;\;\; \vdots \\
\textbf{0}\quad\;\;\;\; \textbf{0}\quad\;\;\;\; \textbf{0}\quad\quad \cdots\quad \textbf{\textit{I}}\quad\;\;\;\; \textbf{0}\\
\textbf{\textit{A}}_{d}\quad \textbf{\textit{A}}_{d-1}\quad \textbf{\textit{A}}_{d-2}\ \,  \cdots\quad \textbf{\textit{A}}_{2}\quad \textbf{\textit{A}}_{1}
\end{matrix}
\right]
\label{eq13}
\end{aligned}
\end{equation}
with I being an identity matrix.

With this relaxation, a particular segment in an utterance is not only related to the preceding segment, but to several preceding segments with a window size of \(d\), which is tunable, and \(d = 1\) falls back to the first-order cases.

\subsubsection{Constructing utterance representations}

The higher-order Koopman operator \(\tilde{\textbf{\textit{A}}}\) can be derived using Equation (11) as follows:
\begin{equation} 
\begin{aligned}
\tilde{\textbf{\textit{A}}} = \tilde{\textbf{\textit{U}}}^{N}_{2} (\tilde{\textbf{\textit{U}}}^{N-1}_{1})^{\dagger}
 \label{eq15}
\end{aligned}
\end{equation}
where ``\(\dagger\)" denotes the pseudoinverse operation.

The dynamic modes and mode amplitudes can then be obtained by calculating the eigenvalues and eigenvectors of \(\tilde{\textbf{\textit{A}}}\). In this paper, the dynamic mode (or eigenvector) that corresponds to the largest dynamic mode amplitude (or eigenvalue) is used as the utterance representation for the corresponding utterance, as it captures the largest-scale dynamic present in the sequence of segments. Algorithm 1 illustrates the overall process. 

\begin{algorithm}
\caption{DMD algorithm for constructing an utterance representation}
\begin{algorithmic}[ht!]
\STATE {\bf Input:} (a) Sequence of segments in an utterance \(\textbf{\textit{U}}^{N}_{1} = [\textbf{\textit{s}}_{1},\ \textbf{\textit{s}}_{2},\ \dots,\ \textbf{\textit{s}}_{N}]\). (b) Order parameter \(d\). 
\STATE {\bf Outputs:} Utterance representation.
\STATE 1: \quad Declare \(\tilde{\textbf{\textit{U}}}^{N}_{1} = [\tilde{\textbf{\textit{s}}}_{1},\ \tilde{\textbf{\textit{s}}}_{2},\ \dots,\ \tilde{\textbf{\textit{s}}}_{N}]\), where \(\tilde{\textbf{\textit{s}}}_{k}\) is given by Equation (12);
\STATE 2: \quad Computing the higher-order Koopman operator \(\tilde{\textbf{\textit{A}}}\):
~\\[-0.2cm]
\begin{equation} \label{eq:P2}
\begin{aligned}
\tilde{\textbf{\textit{A}}} = \tilde{\textbf{\textit{U}}}^{N}_{2} (\tilde{\textbf{\textit{U}}}^{N-1}_{1})^{\dagger} \nonumber
\end{aligned}
\end{equation}
\STATE 3: \quad Performing eigendecomposition on \(\tilde{\textbf{\textit{A}}}\):
~\\[-0.2cm]
\begin{equation} \label{eq:P2}
\begin{aligned}
\left[\textbf{\textit{W}},\ \textbf{\textit{D}}\right] = eig(\tilde{\textbf{\textit{A}}})  \nonumber
\end{aligned}
\end{equation}
where \(\textbf{\textit{D}}\) is a diagonal matrix composed of sorted eigenvalues and the columns of matrix \(\textbf{\textit{W}}\) are the corresponding right eigenvectors;
\STATE 4: The top eigenvector in \(\textbf{\textit{W}}\) that corresponds to the largest eigenvalue in \(\textbf{\textit{D}}\) is selected for constructing the utterance representation.
\label{algo: learning}
\end{algorithmic}
\end{algorithm}

\subsection{Competing Methods}

\subsubsection{P-means}
P-means \cite{ruckle2018concatenated} is a method that concatenates different types of means, also known as power-means \cite{hardy1952inequalities}. The hypothesis is that the average is only one type of order-statistic, and there are several others available, which might add useful information when constructing utterance representations.

\subsubsection{Functionals}
The comparison to functionals is only natural, as it is a common practice within this community. The functionals employed in this work include arithmetic mean, Percentile 1, Percentile 99, and Quartiles 1-3. 

\subsubsection{Discrete Cosine Transform}
The Discrete Cosine Transform (DCT) algorithm is widely used in digital signal processing applications for summarizing or compressing information. In this paper, DCT is applied on the EPs. Taking the BEPs, for example, given an utterance of \(N\) segments \(\textbf{\textit{s}}_{1},\ \textbf{\textit{s}}_{2},\ \dots,\ \textbf{\textit{s}}_{N}\), we stack the sequence of \(M\)-dimensional BEPs into an \(N \times M\) matrix. The DCT algorithm is then applied along the \(M\) columns, respectively. To get a fixed-length utterance representation, we extract and concatenate the first \(K\) DCT coefficients and discard higher-order coefficients, which results in consistent utterance vectors of size \(KM\). For cases where \(N < K\), we pad the utterance with \(K\) - \(N\) zero vectors.

\section{Emotion Corpora}
Two different emotion corpora are used to evaluate the validity and universality of our approach, i.\,e., a Chinese emotional corpus (CASIA) \cite{tao2008design} and an English emotional database (SAVEE) \cite{jackson2014surrey}. All of the emotion categories are selected for each of the two emotion corpora, respectively.

Specifically, the CASIA corpus \cite{tao2008design} contains 9,600 utterances that are simulated by four subjects (two males and two females) in six different emotional states, i.\,e., angry, fear, happy, neutral, sad, and surprise. In our experiments, we only use 7,200 utterances that correspond to 300 linguistically neutral sentences with the same statements. 

The Surrey audio-visual expressed emotion database (SAVEE) \cite{jackson2014surrey} consists of recordings from four male actors in seven different emotions: anger, disgust, fear, happy, sad, surprise, and neutral. Each speaker produced 120 utterances. The sentences were chosen from the standard TIMIT corpus and phonetically-balanced for each emotion. 

\section{Experiments}
\label{sec:experiments}

\subsection{Setup}
According to \cite{provost2013identifying, kim2013emotion}, a speech segment longer than 250 ms contains sufficient emotional information to identify the emotion being expressed in that segment. In our experiment, the size of each speech segment is set to 32 frames, i.e., the total length of a segment is 10 ms \(\times\) 32 + (25 - 10) ms = 335 ms. For the CASIA corpus, the segment hop length is set to 30 ms, while it is set to 10 ms for the SAVEE database. In this way, we collected 418,722 segments for the CASIA corpus and 51,027 segments for the SAVEE database.

For the VGG network, the architecture of the convolutional layers is based on the configurations (i.\,e., configuration E) in the original paper \cite{simonyan2014very}. the only change we made was to the last three FC layers (\(\{128, 32, C\}\) units, respectively, with \(C\) denoting the number of possible emotions). In the training stage, ADAM \cite{kingma2014adam} optimizer with default setting in Tensorflow \cite{abadi2016tensorflow} was used, with an initial learning rate of \(0.001\) and an exponential decay scheme with a rate of \(0.8\) every two epochs. The batch size was set to \(128\). Early stopping with patience of \(3\) epochs was utilized to mitigate an overfitting problem. 

The EPs are generated using leave-one-fold-out ten-fold cross-validation. A \textit{random forest} (RF) with default setting in Scikit-learn \cite{pedregosa2011scikit} was then employed to make the utterance-level decision, where another ten-fold cross-validation is performed. The results are presented in terms of unweighted accuracy (UA) and weighted accuracy (WA), respectively. It is worth noting that the UA and WA are the same for the CASIA corpus as it is perfectly balanced concerning the emotion category.

\subsection{Results and analysis}
Table 1-3 show the results of the experiments performed with P-means, DCT, and DMD on the two stated emotional corpora, respectively. The following can be seen: (1) P-means achieved impressive performance, indicating the importance of the scale information (e.\,g., the average) of an emotional speech utterance. Also, adding higher-order powers was beneficial overall, which corroborated our previous hypothesis. (2) Both of DCT and DMD methods did achieve respectable results, demonstrating that the dynamic information plays an essential role in characterizing the emotional speech as well. (3) Based on the results of the DCT method in Table~\ref{tab:dctresults}, it can be seen that the DCT method needs more components to keep for a relatively large database (CASIA) than a small one (SAVEE), to achieve reasonable performance. (4) Observing the results of the DMD method in Table~\ref{tab:dmdresults}, it is clear that exploiting the higher-order assumption (see Equation 9) is beneficial for the relatively large database (CASIA), since the results are better for adding higher values of the order parameter. (5) P-means outperformed both DCT and DMD-based techniques. We thus posit that the scale information is more critical than dynamics in this task.

\begin{figure*}[thbp]
  \centering
  \includegraphics[width=0.728\linewidth]{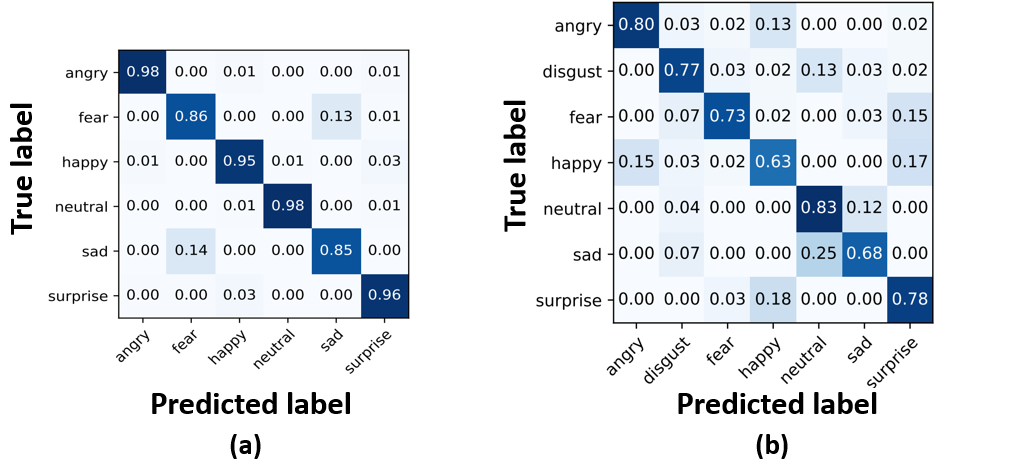}
  \caption{Confusion matrices obtained using the most performant EigenEmo-based utterance representations for (a) CASIA corpus, where the ``DMD$\oplus$AVG \& EEP$\oplus$BEP" method was applied, and (b) SAVEE database, where the ``DMD$\oplus$AVG \& EEP" method was applied, respectively (refer to Table~\ref{tab:comparison}).}
  \label{fig:confusionM}
\end{figure*}

\begin{table}[htbp]
\renewcommand\arraystretch{1.26}
\caption{Results with p-means on the two selected corpora. The Power component is varied between 1, [1-2], [1-3] and [1-6].}
\resizebox{0.96 \linewidth}{!}{%
\begin{tabular}{c|c|c|c|c|c|c|c|c}
\toprule
\multicolumn{1}{c|}{\bf P-means}& \multicolumn{4}{c|}{CASIA} & \multicolumn{4}{c}{SAVEE} \\
\midrule
\multicolumn{1}{c|}{} & \multicolumn{2}{c|}{EEP} & \multicolumn{2}{c|}{BEP} & \multicolumn{2}{c|}{EEP} & \multicolumn{2}{c}{BEP}\\
\midrule
Power(s)  & WA & UA & WA & UA & WA & UA & WA & UA\\
\midrule
$1$ & $92.01$ & $92.01$ & $91.53$ & $91.53$ &  $73.33$ & $71.90$ &  $\bf68.13$ & $\bf65.60$ \\
$[1$-$2]$ & $\bf92.40$ & $\bf92.40$ & $92.11$ & $92.11$ & $73.75$ & $\bf72.38$ &  $65.83$ & $62.98$ \\
$[1$-$3]$ & $92.19$ & $91.19$ & $\bf92.25$ & $\bf92.25$ & $72.50$ & $70.83$ &  $63.96$ & $60.60$ \\
$[1$-$6]$ & $92.17$ & $92.17$ & $92.15$ & $92.15$ & $\bf73.96$ & $\bf72.38$ &  $61.88$ & $58.21$ \\
\bottomrule
\end{tabular}%
}
\label{tab:pmeansresults}
\end{table}
\begin{table}[htbp]
\renewcommand\arraystretch{1.26}
\caption{Results with DCT on the two selected corpora. The number of DCT Components (Cmp.) is varied from \(1\) to \(6\).}
\resizebox{1 \linewidth}{!}{%
\begin{tabular}{c|c|c|c|c|c|c|c|c}
\toprule
\multicolumn{1}{c|}{\bf DCT}& \multicolumn{4}{c|}{CASIA} & \multicolumn{4}{c}{SAVEE} \\
\midrule
\multicolumn{1}{c|}{} & \multicolumn{2}{c|}{EEP} & \multicolumn{2}{c|}{BEP} & \multicolumn{2}{c|}{EEP} & \multicolumn{2}{c}{BEP}\\
\midrule
Cmp. & WA & UA & WA & UA & WA & UA & WA & UA \\
\midrule
$1$ & $89.68$ & $89.68$ & $90.15$ & $90.15$ & $\bf67.08$ & $\bf63.45$ & $\bf62.08$ & $\bf58.57$\\
$2$ & $89.53$ & $89.53$ & $89.78$ & $89.78$ & $61.25$ & $57.38$ & $59.17$ & $54.64$\\
$3$ & $\bf90.04$ & $\bf90.04$ & $90.63$ & $90.63$ & $61.67$ & $57.38$ &$54.17$ & $48.69$\\
$4$ & $89.82$ & $89.82$ & $90.49$ & $90.49$ & $63.33$ & $59.40$ & $54.38$ & $49.17$ \\
$5$ & $89.58$ & $89.58$ & $90.56$ & $90.56$ & $63.54$ & $59.64$ & $52.08$ & $46.07$ \\
$6$ & $89.67$ & $89.67$ & $\bf90.69$ & $\bf90.69$ & $64.38$ & $60.36$ &$51.46$ & $45.24$ \\
\bottomrule
\end{tabular}%
}
\label{tab:dctresults}
\end{table}

\begin{table}[htbp]
\renewcommand\arraystretch{1.26}
\caption{Results with DMD on the two selected corpora. d is the window size as described in Equation (9) and is varied between 1, 2, 3, 6, [1-2], [1-3] and [1-6].}
\resizebox{1 \linewidth}{!}{%
\begin{tabular}{c|c|c|c|c|c|c|c|c}
\toprule
\multicolumn{1}{c|}{\bf DMD}& \multicolumn{4}{c|}{CASIA} & \multicolumn{4}{c}{SAVEE} \\
\midrule
\multicolumn{1}{c|}{} & \multicolumn{2}{c|}{EEP} & \multicolumn{2}{c|}{BEP} & \multicolumn{2}{c|}{EEP} & \multicolumn{2}{c}{BEP}\\
\midrule
d & WA & UA & WA & UA & WA & UA & WA & UA \\
\midrule
$1$ & $90.71$ & $90.71$ & $90.50$ & $90.50$ & $\bf73.96$ & $\bf72.55$ &$\bf63.75$ & $\bf60.83$\\
$2$ & $90.65$ & $90.65$ & $91.03$ & $91.03$ &$73.13$ & $71.48$ &$62.50$ & $59.29$\\
 $3$ & $90.04$ & $90.04$ & $90.25$ & $90.25$ & $72.71$ & $70.95$ &$62.08$ & $58.81$\\
$6$ & $89.88$ & $89.88$ & $89.71$ & $89.71$ & $71.25$ & $69.29$ &$61.67$ & $58.52$\\
$[1$-$2]$ & $90.83$ & $90.83$ & $\bf91.50$ & $\bf91.50$ & $73.54$ & $71.95$ &$61.25$ & $58.10$\\
$[1$-$3]$ & $91.06$ & $91.06$ & $91.29$ & $91.29$ & $72.08$ & $70.71$ &$61.88$ & $58.45$\\
$[1$-$6]$ & $\bf91.33$ & $\bf91.33$ & $91.07$ & $91.07$ & $71.04$ & $69.40$ &$61.04$ & $57.86$\\
\bottomrule
\end{tabular}%
}
\label{tab:dmdresults}
\end{table}

The summary of results is provided in Table~\ref{tab:comparison}, where the best results for each method are provided. It also has an additional result where the most performant EigenEmo-based utterance representations have been concatenated with the averaged EPs. It can be readily seen that this concatenation significantly improved performance, as the resulting representation can now capture both the scale and dynamics of an emotional speech utterance. Figure 2 shows the corresponding confusion matrices.

\begin{table}[htbp]
\renewcommand\arraystretch{1}
\caption{Comparison of methods on the two selected corpora. ``\(\oplus\)" means features are concatenated.}
\resizebox{1 \linewidth}{!}{%
\begin{tabular}{c|c|c|c|c|c}
\toprule
\multicolumn{2}{c|}{}& \multicolumn{2}{c|}{CASIA} & \multicolumn{2}{c}{SAVEE} \\
\midrule
\bf Method & \bf EP Type & WA & UA & WA & UA \\
\midrule
P-means & EEP & $92.40$ & $92.40$ & $73.96$ & $72.38$\\
P-means & BEP & $92.25$ & $92.25$ & $68.13$ & $65.60$\\
P-means & EEP$\oplus$BEP & $92.33$ & $92.33$ & $73.13$ & $71.69$\\
\midrule
DCT & EEP & $90.04$ & $90.04$ & $67.08$ & $63.45$ \\
DCT & BEP & $90.69$ & $90.69$ & $62.08$ & $58.57$ \\
DCT & EEP$\oplus$BEP & $91.10$ & $91.10$ & $65.83$ & $63.67$  \\
\midrule
Functionals & EEP & $92.53$ & $92.53$ & $74.15$ & $73.03$ \\
Functionals & BEP & $92.36$ & $92.36$ & $64.79$ & $62.52$ \\
Functionals & EEP$\oplus$BEP & $92.47$ & $92.47$ & $73.96$ & $72.55$ \\
\midrule
\midrule
DMD & EEP & $91.33$ & $91.33$ & $73.96$ & $72.55$ \\
DMD & BEP & $91.50$ & $91.50$ & $63.75$ & $60.83$ \\
DMD & EEP$\oplus$BEP & $92.07$ & $92.07$ & $71.46$ & $70.60$\\
\midrule
DMD$\oplus$AVG & EEP & $92.04$ & $92.04$ & $\bf75.83$ & $\bf74.76$ \\
DMD$\oplus$AVG & BEP & $92.50$ & $92.50$ & $68.13$ & $67.33$ \\
DMD$\oplus$AVG & EEP$\oplus$BEP & \bf93.28 & \bf93.28 & $73.33$ & $71.38$\\
\bottomrule
\end{tabular}%
}
\label{tab:comparison}
\end{table}

\section{Conclusions}
In this paper, we proposed a novel method to construct utterance representation for speech emotion classification by exploiting the dynamic properties of the emotion profiles generated by a VGG network. We do this using a spectral decomposition method rooted in fluid-dynamics, known as Dynamic Mode Decomposition. Empirical validation of the proposed method on the CASIA corpus and the SAVEE database shows promising results. Since we herein blindly used all segments to train the segment-level classifier, it is anticipated with proper segment selection strategy, better results are expected.

\newpage
\newpage
\bibliographystyle{IEEEtran}
\begin{footnotesize}
\bibliography{mybib}
\end{footnotesize}

\end{document}